# Hysteretic phenomena in graphene's conductivity

A.I.Kurchak[1], A.N.Morozovska[2], M.V.Strikha[1]
[1]V.Lashkariov Institute of Semiconductor Physics, NAS, 41, pr. Nauky, 03028 Kyiv, Ukraine;  [2]Institute of Physics, NAS, 46, pr. Nauky, 03028 Kyiv, Ukraine,

This text is a summary of the article "Rival mechanisms of hysteresis in the resistivity of graphene channel" by A.I.Kurchak, A.N.Morozovska, M.V.Strikha (see Ukr. J. Phys. **58,** 473 (2013), http://ujp.bitp.kiev.ua/index.php?item=j&id=163 ).

Despite a remarkable symmetry of the graphene's band structure, a hysteretic behavior of the graphene's resistivity with gate voltage $V_g$ onward and backward sweeps is often observed in ambient conditions. The mechanism of this hysteresis can be different, depending on the charge transfer between graphene and its environment. Two types of hysteresis directions are observed experimentally [1]: a direct one and an inverse one (the last is also defined as an 'antihysteresis' for the systems on ferroelectric substrate). This hysteresis in the graphene-on-ferroelectric systems had already enabled the construction of the robust elements of non-volatile memory of new generation [2, 3]. These elements work for more than $10^5$ switches and preserve information for more than 1000 s. Such systems can be characterized theoretically by the ultrafast rate of switching (~ 10-100 fs) (for details see rev. [4]).

The direct and opposite hysteresis can be attributed to two rival mechanisms: capacitive coupling and charge carriers trapping from graphene, respectively. A quantitative model for these competing hysteretic mechanisms in graphene channel resistivity on a substrate of different nature – a direct one (caused by adsorbates with dipole moment on surface, e.g. water, as it is presented in fig.1a and an inverse one (caused by capture of free carriers onto localized states on graphene-substrate interface) was proposed in [5, 6].

This model leads in the first case to non-zero concentration of holes in graphene, at zero $V_g$, equal to

$$n_{p0} = P_0 / e \qquad (1)$$

where $P_0$ is the absorbed molecules layer polarization, i.e. the electro-neutrality point on $V_g$ scale is shifted to positive voltages. Therefore the linear dependence $n(V_g)$ is shifted initially to smaller values of electron concentrations (and greater values of holes concentrations). Further increase of $V_g$ ruins finally the polarization, as it is presented in fig.1b, and for the backward sweep of $V_g$ the dependence $n(V_g)$ is governed by the fleet capacitor correlation, with electro-neutrality point at zero $V_g$. Therefore the hysteresis in the dependence of conductivity σ($V_g$) occurs (fig.1c). However, it $V_g$ sweeps rate is slow enough, the spontaneous polarization on graphene's surface restores with time as:

$$P(t) = P_0(1 - \exp(-t/\tau)) \qquad (2)$$

If the time, when $V_g$ reaches electro-neutrality, is of $\tau$ constant order or greater, the two electro-neutrality points in fig.1c trends to one and the hysteresis vanishes.

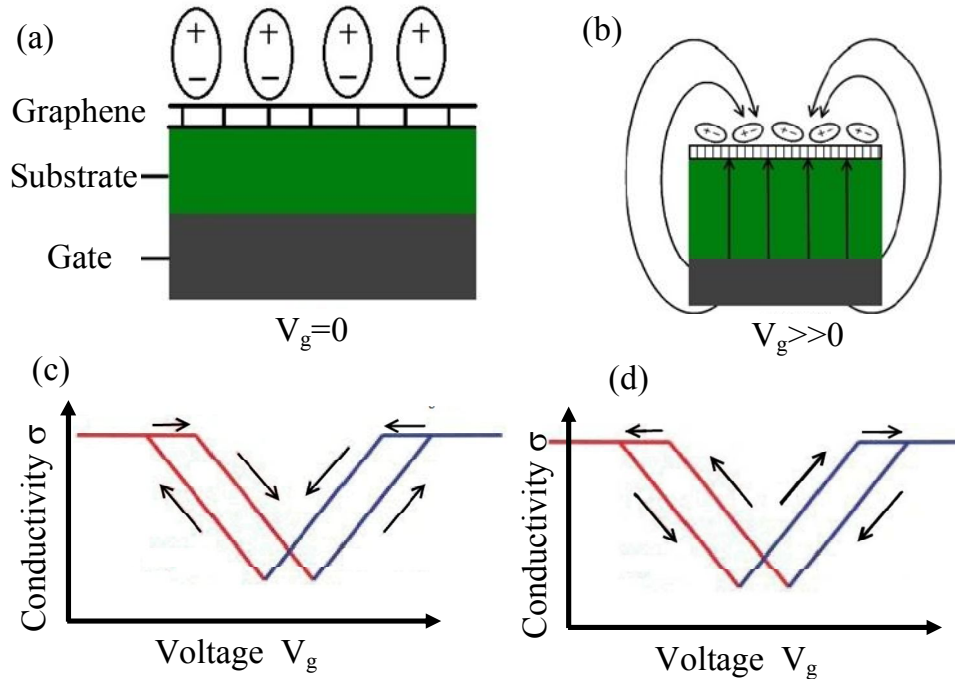

Fig.1. (a) Graphene with adsorbed dipoles at the surface. (b) Dipoles reorientation under high gate voltages. Direct (c) and inverse (d) conductivity hysteresis σ($V_g$).

The second channel was examined in [5]. If there are localized interface states with $E_T$ energy and $n_T$ concentration, electrons from graphene are captured onto them when Fermi energy reaches $E_T$ on $V_g$ onward sweep, and $n(V_g)$ shifts in $n_T$ value lower. The characteristic lifetime of electrons on interface states $\tau_c$ is rather long [5], therefore this shift remains in $n(V_g)$ dependence on $V_g$ backward sweep as well, and this causes the inverse hysteretic loop, presented in fig.1d.

Mention, that due to hierarchy of times $\tau_c \gg \tau$ the realistic sweep rate variation vanishes the direct hysteresis primarily. This was observed experimentally in [1], and $\tau = 3{,}4$ s can be obtained from it's data [6]. This time is many orders of value longer than the relaxation time of free dipoles due to strong chemical bonds of adsorbates to graphene's surface. Another experimental result of [1] that can be explained within our model is that hysteresis could be swept with the decrease of $V_g$ sweep rate from direct to inverse one for low temperatures, but for room temperatures it was inverse one only. Although chemical nature of adsorbate (water) was the same, $\tau$ for ice is many orders greater, than for liquid, therefore no direct hysteresis could be observed for room temperature due to rapid polarization renewal.

Thus, possible discrimination of these two channels by variation of the gate voltage sweep rate was proposed: the slow sweep suppresses the direct channel and does not influence greatly for the realistic sweep rate the opposite one. The theory's predictions are in a qualitative agreement with experimental data and can be useful for understanding of the graphene's structures operation in ambient conditions.

This work was supported by Fundamental Research Fund of Ukraine, NAS of Ukraine and STCU.